\documentclass[12pt]{article}
\usepackage{geometry}
\usepackage[round]{natbib}
\usepackage{graphicx}
\usepackage[T1]{fontenc}
\usepackage[utf8]{inputenc}
\usepackage{authblk}
\usepackage[running]{lineno}
\usepackage{setspace}

\usepackage[dvipsnames]{xcolor}
\usepackage[colorinlistoftodos,textwidth=2.7cm]{todonotes}
\usepackage{lscape}
\usepackage{longtable}
\usepackage{multirow}
\usepackage{array}
\usepackage{booktabs} 
\usepackage{ulem}
\usepackage{url}
\makeatletter
\g@addto@macro{\UrlBreaks}{\UrlOrds}
\makeatother

\newcommand{\Rpkg}[1]{\texttt{#1}}

\definecolor{red1}{RGB}{228,26,28}
\definecolor{red2}{RGB}{251,180,174}
\definecolor{blue1}{RGB}{55,126,184}
\definecolor{blue2}{RGB}{179,205,227}
\definecolor{green1}{RGB}{77,175,74}
\definecolor{green2}{RGB}{204,235,197}
\definecolor{purple1}{RGB}{152,78,163}
\definecolor{purple2}{RGB}{222,203,228}
\definecolor{orange1}{RGB}{255,127,0}
\definecolor{orange2}{RGB}{254,217,166}
\definecolor{mustard1}{RGB}{191,129,45}
\definecolor{mustard2}{RGB}{254,217,118}
\definecolor{pink1}{RGB}{247,129,191}
\definecolor{pink2}{RGB}{253,218,236}
\definecolor{blueish}{RGB}{122,200,246}
\definecolor{greenish}{RGB}{161,217,155}

\title{Navigating through the R packages for movement}
\author[a]{Rocio Joo\thanks{rocio.joo@ufl.edu}}
\author[a]{Matthew E. Boone}
\author[b]{Thomas A. Clay}
\author[b]{Samantha C. Patrick}
\author[c]{Susana Clusella-Trullas}
\author[a]{Mathieu Basille}
\affil[a]{Department of Wildlife Ecology and Conservation, Fort Lauderdale Research and Education Center, University of Florida, Fort Lauderdale, FL, USA}
\affil[b]{School of Environmental Sciences, University of Liverpool, Liverpool, L69 3GP, UK}
\affil[c]{Department of Botany and Zoology and Centre for Invasion Biology, Stellenbosch University, Stellenbosch, South Africa}

\date{*Corresponding author: rocio.joo@ufl.edu}

\begin{document}
\maketitle

\section*{Summary}
\begin{enumerate}
  \item The advent of miniaturized biologging devices has provided ecologists with unprecedented opportunities to record animal movement across scales, and led to the collection of ever-increasing quantities of tracking data. In parallel, sophisticated tools have been developed to process, visualize and analyze tracking data, however many of these tools have proliferated in isolation, making it challenging for users to select the most appropriate method for the question in hand. Indeed, within the R software alone, we listed 58 packages created to deal with tracking data or `tracking packages'.
  \item Here we reviewed and described each tracking package based on a workflow centered around tracking data (i.e. spatio-temporal locations $(x,y,t)$), broken down into three stages: pre-processing, post-processing and analysis, the latter consisting of data visualization, track description, path reconstruction, behavioral pattern identification, space use characterization, trajectory simulation and others. 
  \item Supporting documentation is key to render a package accessible for users. Based on a user survey, we reviewed the quality of packages' documentation, and identified 11 packages with good or excellent documentation.
  \item Links between packages were assessed through a network graph analysis. Although a large group of packages showed some degree of connectivity (either depending on functions or suggesting the use of another tracking package), one third of the packages worked in isolation, reflecting a fragmentation in the R movement-ecology programming community.
  \item  Finally, we provide recommendations for users when choosing packages, and for developers to maximize the usefulness of their contribution and strengthen the links within the programming community.
   \end{enumerate}

\section*{Keywords}
biologging, movement ecology, R project for statistical computing, spatial, tracking data

\section*{A Movement Ecology background}

Animal movement plays a crucial role in ecological and evolutionary processes, from the individual to ecosystem level \citep{dingle_migration:_1996, clobert_dispersal_2001, nathan_movement_2008}. However, studying animal movement has presented challenges to researchers, as individuals are often difficult to follow for extended time periods and over large distances. Over recent decades, decreases in the size and cost of animal-borne sensors or biologging devices have led to an exponential increase in their use. This has substantially improved our understanding of how and why animals move \citep{nathan_movement_2008, kays_terrestrial_2015, hussey_aquatic_2015}. Technological advancements have also enabled a wide range of sensors to be used by ecologists, which can be integrated to remotely record a suite of metrics, including longitude and latitude  $(x, y)$, altitude or depth  $(z)$, acceleration, as well as in-situ environmental conditions  \citep{wilson_prying_2008, cagnacci_animal_2010, wilmers_golden_2015}. From these multiple sensors, fine-scale behaviors and physiological states can be inferred \citep{rutz_new_2009, halsey_accelerometry_2009}. 

The increase in quantity and complexity of biologging data requires appropriate analytical and software tools that aid processing and interpretation of data. Those tools should be sound and transparent to allow for reproducibility of results and computation time optimization \citep{Urbano2010, reichman_challenges_2011, lowndes_our_2017}. 
Mainly in the last decade, many of these tools have been made available for the scientific community in the form of packages for the R software \citep{R2018}, which has facilitated their widespread use and contributed to make R the most dynamic programming platform in ecology. However, in order to identify the most appropriate function in R for a particular analysis, ecologists  have to review and evaluate multiple functions within and between packages. 

The aim of this study is to review the packages created to process or analyze a specific type of movement data: tracking data. 
Movement of an organism is defined as a change in the geographic location of an individual in time, so movement data can be defined by a space and a time component. Tracking data are composed by at least 2-dimensional coordinates $(x,y)$ and a time index $(t)$, and can be seen as the geometric representation (the trajectory) of an individual's path. 
The packages reviewed here, henceforth called tracking packages, are those explicitly developed to either create, transform or analyze tracking data (i.e. $(x,y,t)$), allowing a full workflow from raw data from biologging devices to final analytical outcome. For instance, a package that would use accelerometer, gyroscope and magnetometer data to reconstruct an animal's trajectory via dead-reckoning, thus transforming those data into an $(x,y,t)$ format, would fit into the definition. However, a package analyzing accelerometry series to detect changes in behavior would not fit.

Here, we present a workflow for the study of tracking data (Fig. $1$) and review packages that are designed for tracking data, including their role in data processing and analysis. The workflow is composed of three stages: pre-processing, post-processing and data analysis. Data pre-processing is the process by which data are transformed into the $(x,y,t)$ format, and it would be necessary in cases where biologging devices do not provide raw data in the form of tracking data, e.g. for most geolocators or Global Location Sensors (GLS), only light intensity is provided. Tracking data may not be immediately usable, e.g. errors or outliers need to be identified, or other second or third order variables need to be derived for the dataset to be ready for analysis; we defined this stage of data processing as post-processing. Finally, the last stage of analysis can be divided into data visualization, track description, path reconstruction, behavioral pattern identification, space use characterization, trajectory simulation and others (e.g. population parameter estimation, interaction between individuals). In each of these subsections, we describe the tools provided by tracking packages to achieve these goals. When necessary, we also provided a short description of the biologging devices and the data they collect, since not all readers are familiar with every type of device. An additional subsection briefly describes some R packages that do not deal with tracking data (as defined above), but were developed to process and analyze data from biologging devices such as accelerometers and time-depth recorders. 

Since the documentation provided in conjunction with the packages are key for rendering them accessible for users, we also review supporting documentation and, based on a survey, summarize packages based on the clarity of their documentation. The links between packages, showing how much they rely on each other and the compatibility between them, are also assessed. 

This review is aimed at movement ecologists, whether they are potential users or developers of R packages. This study aims to provide users with criteria through which they can select packages for specific analyses, and offers developers recommendations to maximize the utility of packages and strengthen the links within the R community. 

\section*{Data sources}

Multiple sources were used to identify tracking packages; mainly, 1) the spatio-temporal task view on the Comprehensive R Archive Network (CRAN) repository (\url{https://cran.r-project.org/web/views/SpatioTemporal.html}, 2) an updated list of this task on GitHub (\url{https://bit.ly/2CWoSD6}), 3) packages suggested in the description files of other packages, 4) Google search engine and 5) e-mail/Twitter exchanges with ecologists. For the Google search, search terms were (trajectory OR movement OR spatiotemporal) AND package AND R. The combined use of these sources provided a large list of packages, from which we selected only the ones that matched the definition of a tracking package stated in the previous section (\emph{A Movement Ecology background}).

The package search and information gathering was done between March and August 2018. Tracking packages that were either removed from CRAN or described as in a `very early version' on their GitHub repositories were discarded.  Information on package documentation was extracted as follows. Standard documentation was categorized as existing if it was available when installing the package. A vignette had to be visible from the main page of the repository or visible as an output of \textit{help(package)}. A peer-reviewed article had to be either mentioned on the main page of the repository, the vignette or in the citation of the package. 

It should be noticed that between the period of information gathering and the time of publication of this work, new packages may have been published, and new versions of the reviewed packages containing additional functions could have been released. Information on the reviewed version of each package, links to each package repository along with a summary of their main characteristics are included in the Zenodo repository (\url{https://doi.org/10.5281/zenodo.3483853}). In this work, package citation refers strictly to the output of \textit{citation(package)} in R by the time of writing the manuscript and the version cited in the reference may not match the studied version.

\section*{The R packages}

Fifty-eight packages were found to assist with processing and analysis of tracking data (Fig. $1$). Some R packages have been developed to tackle several of these stages of data processing and analysis, while others focus on only one, as shown in Table \ref{table:PurposeTable}. To identify and classify packages functions for each specific stage, our main support was the standard documentation of the packages, complemented with the additional sources of information described in section \emph{Data sources} above.

When appropriate, the type of biologging devices from which the tracking data originate is described in the text, so that readers that are not familiar with these devices have a basic idea of the advantages and limitations of the devices, and why some packages focus on specific issues related to them. The description of the tracking packages 
also includes information on the year each package was publicly available (Fig. $2$), the main repository where the package is stored and whether it is actively maintained (hereafter referred to as `active'). The official repository for R packages is the CRAN repository. CRAN enforces technical consistency, with a set of rules such as the inclusion of ownership information, cross-platform portable code (i.e. to work with Windows, Mac OS and UNIX platforms), minimum and maximum sizes for package components, among others. The majority of the packages reviewed in this manuscript are on CRAN; the remainder are mostly on GitHub or other repositories (e.g. R-Forge or independent websites). Regarding package maintenance, we consider that a package hosted on GitHub is actively maintained if a `commit' (i.e. a contribution) has been made in the last year, and for other packages (if they are not also on GitHub), that the most recent version of the package is no older than one year (analysis conducted in August 2018). 

\subsection*{Pre-processing} 

Pre-processing is required when raw biologging data are not in a tracking data format. The methods used for pre-processing depend heavily on the type of biologging device used. Among the tracking packages, 6 are focused on GLS, one on radio telemetry, and two use accelerometry and magnetometry data.

\subsubsection*{GLS data pre-processing} 

GLS are electronic archival tracking devices which record ambient light intensity and elapsed time. The timings of sunset and sunrise are estimated, latitude is calculated from day length, and longitude from the time of local midday relative to Greenwich Mean time \citep{Afanasyev2004}. GLS can record data for several years and their small size and low mass ($<$1 g) make them suitable for studying long-distance movements in a wide range of species. Several methodologies have been developed to reduce errors in geographic locations generated from the light data, which is reflected by the large number of packages for pre-processing GLS data. We classified these methods in three categories: threshold, template-fitting and twilight-free.

\begin{itemize}
        \item Threshold methods. Threshold levels of solar irradiance, which are arbritrarily chosen, are used to identify the timing of sunrise and sunset. The packages that use threshold methods are \Rpkg{GeoLight} (2012, CRAN, inactive) \citep{Lisovski2012} and \Rpkg{probGLS} (2016, GitHub, inactive) \citep{RprobGLS}. \Rpkg{GeoLight} uses astronomical equations from \cite{Montenbruck2013} to derive locations from timings of sunrise and sunset, and from sun elevation angles. \Rpkg{probGLS} implements a probabilistic method that takes into account uncertainty in sun elevation angle and twilight events to estimate locations. Starting with the first known location (where the individual was tagged), it estimates the location of the subsequent twilight event which is replicated several times adding an error term; it then computes probabilities for each location based on the plausibility of the estimated speed or on environmental conditions (e.g. sea surface temperature SST) \citep{Merkel2016}. 
        \item Template-fitting methods. The observed light irradiance levels for each twilight are modeled as a function of theoretical light levels (i.e. the template). Then, parameters from the model (e.g. a slope in a linear regression) are used to estimate the locations. The formulation of the model and the parameters used for location estimation vary from method to method \citep{Ekstrom2004}. The packages that use template-fitting methods are \Rpkg{FLightR} (2015, CRAN, active) \citep{RFLightR}, \Rpkg{trackit} (2012, GitHub, active) \citep{Rtrackit} and \Rpkg{tripEstimation} (2007, GitHub, inactive) \citep{Sumner2009,RtripEstimation}. \Rpkg{FLightR} was specifically developed for avian movement. In its state-space modeling framework \citep{Patterson2008}, the locations are hidden states and the observation model is a physical model of light level changes as a function of geographic location and time. A detailed description of the model and the package functions can be found in \cite{Rakhimberdiev2015} and \cite{Rakhimberdiev2017}, respectively. \Rpkg{trackit} was developed mainly for fish movement and light intensity around sunrise and sunset are used as inputs in a state-space model that includes solar altitude and SST as covariates \citep{Lam2010}. \Rpkg{tripEstimation} was developed for marine organisms. It uses a Bayesian approach modeling light level as a function of sun elevation at each plausible location, prior knowledge of the animal's movement, and complementary environmental information (e.g. SST, depth of the water column) \citep{Sumner2009}. Although \Rpkg{tripEstimation} is still available on CRAN, it is indicated in its GitHub repository that the package was deprecated in favor of \Rpkg{SGAT} \citep{Sumner2009,Lisovski2012}, which contains functions to implement both threshold and template-fitting methods (note that the authors of \Rpkg{tripEstimation} are also the main authors of \Rpkg{SGAT} and \Rpkg{GeoLight}, and that the references to cite the packages are the same). For this reason, we consider both \Rpkg{tripEstimation} and \Rpkg{SGAT} as one. Auxiliary packages also exist to detect the timing of twilight periods from light data from GLS devices (e.g. \Rpkg{TwGeos} \citep{RTwGeos} and \Rpkg{BAStag} \citep{RBAStag}). The estimated twilight periods can be later used as inputs in the above mentioned packages for location estimation.
        \item Twilight-free methods. It is possible to estimate locations without depending on the identification of twilight events. 
        \Rpkg{TwilightFree} (2017, GitHub, active) \citep{RTwilightFree} uses a Hidden Markov Model (HMM) where the hidden states are the daily geographic locations (the spatial domain is discretized as gridded cells) and the observed variable is the observed pattern of light and dark over the day \citep{Bindoff2017}. SST and land/sea marks can be used as covariates. Parameter estimation is performed using functions from the \Rpkg{SGAT} package.
\end{itemize}

\subsubsection*{Radio tagging data pre-processing}

Radio tagging \citep{Kenward2000} involves the attachment of a radio transmitter to an animal. The radio signals transmitted (typically Very High Frequency VHF or Ultra High Frequency UHF) are picked up by an antenna and transformed into a beeping sound by a receiver. As the receiver gets closer to the transmitter, the beeps get louder. Location can then be estimated either by triangulation or with a method called homing, where the researcher moves towards the loudest beeps until the animal has been located. RFID (radio-frequency identification data) tags can also be used to record when an individual passes close to a receiver without the need to search for a signal. 

\Rpkg{telemetr} (2012, GitHub, inactive) \citep{Rtelemetr} implements several triangulation methods as well as a maximum likelihood procedure to estimate locations from bearing data (triangulation information). Since there are no references to the methods in the package documentation, it is aimed at users that are already familiar with the methods.

\subsubsection*{Dead-reckoning using accelerometry and magnetometry data} 
 
High-frequency (e.g. $>$10 Hz) tri-axial accelerometers measure both static (gravitational) and dynamic body acceleration (DBA). The static component is typically derived using a sliding average over short time windows of a few seconds on each axis \citep{Shepard2008}. The static component enables determining the animal's body posture. The dynamic component is calculated by subtracting the static acceleration from the raw acceleration on each axis, and provides a measure of the animal's movement or velocity as a result of body motion. When coupled with time activity budgets and validated with empirical measurements of metabolic rate, the overall DBA can be used to estimate the animal's energy expenditure \citep{Wilson2019}. High-frequency tri-axial magnetometers measure the geomagnetic field strength in the three axes and provide a measure of 3D orientation for dead-reckoning \citep[e.g.][]{Bidder2015} and for behavior identification \citep{Williams2017}. 

The combined use of magnetometer and accelerometer data, especially as provided by modern inertial measurement units, which solve the problem of temporal synchronization among different sensors, and optionally gyroscopes and speed sensors, allows to reconstruct sub-second fine scale movement paths using the dead-reckoning (DR) technique \citep{Wilson1991,Bidder2015}. Given an initial known location (e.g. tagging or release location), the DR method uses speed and direction movement parameters derived from accelerometer, magnetometers and sometimes additional sensors, to reconstruct the movement path from one location to the next. Specifically, DBA derived from accelerometers can provide a useful metric of speed for terrestrial individuals \citep{Bidder2012}, though in aerial/aquatic media it may be better to use a speed sensor. Magnetometers---after appropriate calibration and correction for other sources of magnetism \citep{Bidder2015}, and in combination with accelerometers and gyroscopes when available---provide fine-scale measures of heading and direction. However, as DR is based on vectorial calculations, it accumulates errors over time, further compounded in the presence of passive movements caused by currents and drifts. Independent locations, typically collected by a GPS recording at lower frequency than the accelerometers and magnetometers, are required to correct for these errors \citep{Bidder2015,Liu2015}; see section \emph{Path reconstruction} below for further details. Furthermore, the exact mathematical formulas for DR differ in the literature, and most of them do not account for 3D movement (see \cite{Benhamou2018} for a comparison of movement properties in 2D and 3D). A discussion on DR per se is out of the scope of this work, but we advise users to understand the methods behind the packages performing DR before using them. \Rpkg{animalTrack} (2013, CRAN, inactive) \citep{RanimalTrack} and \Rpkg{TrackReconstruction} (2014, CRAN, inactive) \citep{RTrackReconstruction} implement DR to obtain tracks, though use different methods. While \Rpkg{TrackReconstruction} refers to \cite{Wilson2007} for DR, \Rpkg{animalTrack} cites \cite{Bowditch1995}.

\subsection*{Post-processing}

Post-processing of tracking data comprises data cleaning (e.g. identification of outliers or errors), compressing (i.e. reducing data resolution which is sometimes called resampling) and computation of metrics based on tracking data, which are useful for posterior analyses.

\subsubsection*{Data cleaning}

\Rpkg{argosfilter} (2007, CRAN, inactive) \citep{Rargosfilter} and \Rpkg{SDLfilter} (2014, CRAN, active) \citep{Shimada2012,Shimada2016} implement functions to filter implausible platform terminal transmitter (PTT) locations. Platform terminal (also known as Argos) transmitters send signals to polar-orbital Argos satellites, which geographically locate the source of the data. They preserve battery life by only needing to transmit signals (rather than receiving), leading them to be used for tracking of large-scale migrations, particularly marine mammals and turtles. When the tracked animals are under water, the chances of a satellite receiving PTT signals decrease, so fewer locations can be estimated, and they are likely estimated with fewer satellites, so their accuracy also diminishes. PTTs are particularly useful for individuals that cannot be recaptured, and hence a device recovered. Along with locations, Argos provide accuracy classes (1, 2, 3, 0, A, B, Z) which are associated with different degrees of spatial error \citep{Costa2010}. \Rpkg{argosfilter}'s algorithm is described in \cite{Freitas2007}. It essentially removes records where a location was not estimated as well as locations that required unrealistic travel speeds. \Rpkg{SDLfilter} allow the removal of duplicates, locations estimated with a low number of satellites, biologically unrealistic locations based on speed thresholds or turning angles and locations above high tide lines. The filtering methods are described in \cite{Shimada2012,Shimada2016}, and they are also adapted to GPS data. GPS loggers are perhaps the most widely used type of biologging device. Location information from GPS can be downloaded directly without any post processing. GPS receivers collect but do not transmit information, and infer their own location based on the location of GPS satellites and the time of transmission. Four or more satellites should be visible to the receiver to obtain an accurate result ($<100$ m; able to reach $6$ m in some cases) \citep{Tomkiewicz2010}, so when less satellites are visible, location accuracy can be reduced. 

Other packages with functions for cleaning tracking data are \Rpkg{T-LoCoH} (2013, R-forge, active) \citep{Rtlocoh}, \Rpkg{TrajDataMining} (2017, CRAN, active) \citep{RTrajDataMining} and \Rpkg{trip} (2006, CRAN, active) \citep{Rtrip}. They can be used for any tracking data and also contain functions to remove duplicates or records with unrealistically high speeds.

\subsubsection*{Data compression}

Rediscretization or getting data to equal step lengths can be achieved with \Rpkg{adehabitatLT} (2010, CRAN, active) \citep{Calenge2006}, \Rpkg{trajectories} (2014, CRAN, active) \citep{Rtrajectories} or \Rpkg{trajr} (2018, CRAN, active) \citep{McLean2018}. Regular time-step interpolation can be performed using \Rpkg{adehabitatLT}, \Rpkg{amt} (2016, CRAN, active) \citep{Ramt} or \Rpkg{trajectories}. Other compression methods include Douglas-Peucker (\Rpkg{TrajDataMining} and \Rpkg{trajectories}), opening window (\Rpkg{TrajDataMining}) or Savitzky-Golay (\Rpkg{trajr}). For a brief review on compression methods, see \cite{Meratnia2004}.

\Rpkg{rsMove} (2017, CRAN, active) \citep{RrsMove} provides functions to explore and transform tracking data for a posterior linkage with remote sensing data. Location fixes are transformed into pixels and grouped into regions. The spatial or temporal resolution of the tracking data can be changed to match the resolution of the remote sensing data. 

\subsubsection*{Computation of metrics}

Some packages automatically derive second or third order movement variables (e.g. distance and angles between consecutive fixes) when transforming the tracking data into the package's data class (most packages define their own data classes, see file in Zenodo, \url{https://doi.org/10.5281/zenodo.3483853}). These packages are \Rpkg{adehabitatLT}, \Rpkg{momentuHMM} (2017, CRAN, active) \citep{McClintock2018}, \Rpkg{moveHMM} (2015, CRAN, active) \citep{Michelot2016}, \Rpkg{rhr} (2014, GitHub, inactive) \citep{Rrhr} and \Rpkg{trajectories}. \Rpkg{bcpa} has a function to compute speeds, step lengths, orientations and other attributes from a track. \Rpkg{amt}, \Rpkg{move} (2012, CRAN, active) \citep{Rmove}, \Rpkg{segclust2d} (2018, CRAN, active) \citep{Rsegclust2d}, \Rpkg{trajr} and \Rpkg{trip} also contain functions for computing those metrics, but the user needs to specify which ones they need to compute. 

\Rpkg{feedr} (2016, GitHub, active) \citep{Rfeedr} works specifically with RFID data (described in subsection \emph{Radio tagging data pre-processing} above). Raw RFID data typically contain an individual line of data for each read event made by each RFID logger. \Rpkg{feedr} contains functions to read raw data from several RFID loggers, and to transform the data of logger detection into movement data for each individual, computing statistics such as the time of arrival and departure from each logger station, and how much time was spent near a station at each visitation. 

\Rpkg{VTrack} (2015, CRAN, active) \citep{Campbell2012} handles acoustic telemetry data. Acoustic telemetry uses high frequency sound (between 30 and 300 kHz) to transmit information through water. Tags (transmitters) emit a pulse of sound, which is detected by a hydrophone (or an array of hydrophones) with an acoustic receiver. The distance at which a transmitter can be detected depends on the power and frequency of the tag, and the characteristics of the surrounding environment (e.g. background noise, water turbidity and temperature) \citep{Decelles2014}. \Rpkg{VTrack} was created to deal with VEMCO\copyright  data, which has a similar structure than RFID; it is composed of transmitter ID, receiver ID, datetime stamps and the location of receiver. Like \Rpkg{feedr} for RFID, \Rpkg{VTrack} can compute statistics such as the time of arrival and departure from each receiver, and how much time was spent near a receiver at each visitation.

\subsection*{Visualization}

In this section, we focus on the packages mainly developed for visualization purposes. Those are \Rpkg{anipaths} (2017, CRAN, active) \citep{Ranipaths} and \Rpkg{moveVis} (2017, CRAN, active) \citep{RmoveVis}. 

They were both conceived for producing animations of tracks. \Rpkg{anipaths} relies on the \Rpkg{animation} package \citep{Xie2013,Ranimation}. Users can specify time-steps and seconds per frame for animation, add a background map (e.g. Google Maps) and an individual-level covariate (e.g. migrant, stationary), among others. Consecutive fixes are joined via a spline-based interpolation and a confidence interval for the interpolation of the path for animation can be shown. 

\Rpkg{moveVis} is based on a \Rpkg{ggplot2} \citep{Wickham2016} plotting architecture and works with \Rpkg{move} data class objects. Users can choose between `true time' which displays the animation respecting the timestamps provided, or `simple' animations where time is not taken into account and all individuals are displayed together as if their tracks started at time 0. Consecutive fixes are joined via linear interpolation. As in \Rpkg{anipaths}, users can specify the number of frames per second and personalize the background map. Statistics related to the background layer (e.g. temperature, land cover) can also be shown as animated lines or bar plots. For both packages, animations can be saved in many different formats such as mpeg, mp4 and gif. 

\subsection*{Track description}

\Rpkg{amt}, \Rpkg{movementAnalysis} (2013, GitHub, inactive) \citep{RmovementAnalysis} and \Rpkg{trajr} compute summary metrics of tracks, such as total distance covered, straightness index and sinuosity. It should be noted that \Rpkg{movementAnalysis} depends on \Rpkg{adehabitat}, which was officially removed from CRAN in 2018, as it was superseeded by \Rpkg{adehabitatLT}, \Rpkg{adehabitatHR} (2010, CRAN, active) \citep{Calenge2006} and \Rpkg{adehabitatMA} \citep{Calenge2006} in 2010. 

\Rpkg{trackeR} (2015, CRAN, active) \citep{Frick2017} was created to analyze running, cycling and swimming data from GPS-tracking devices for humans. \Rpkg{trackeR} computes metrics summarizing movement effort during each track (or workout effort per session). Those metrics include total distance covered, total duration, time spent moving, work to rest ratio, averages of speed, pace and heart rate. 

\subsection*{Path reconstruction}

Whether it is for the purposes of correcting for sampling errors, or obtaining finer data resolutions or regular time steps, path reconstruction is a common goal in movement analysis. Here, we mention methods available, however, before choosing a method, users should be aware that every method is constructed under unique movement assumptions (either inherent to the mathematical model or constructed for a particular species or type of data), and users should refer to the literature on the methods first. Packages available for path reconstruction are \Rpkg{HMMoce} (2017, CRAN, active) \citep{Braun2017}, \Rpkg{kftrack} (2011, GitHub, active) \citep{Rkftrack}, \Rpkg{ukfsst/kfsst} (2012, GitHub, active) \citep{Rukfsst}, \Rpkg{argosTrack} (2014, GitHub, active) \citep{RargosTrack,Albertsen2015}, \Rpkg{bsam} (2016, CRAN, active) \citep{Jonsen2005,Jonsen2016}, \Rpkg{BayesianAnimalTracker} (2014, CRAN, inactive) \citep{RBayesianAnimalTracker}, \Rpkg{TrackReconstruction}, \Rpkg{crawl} (2008, CRAN, active) \citep{Rcrawl,Johnson2008}, \Rpkg{ctmcmove} (2015, CRAN, active) \citep{Rctmcmove} and \Rpkg{ctmm} (2015, CRAN, active) \citep{Rctmm}. While the first three focus on GLS data, \Rpkg{bsam} is intended for PTT data, \Rpkg{BayesianAnimalTracker} and \Rpkg{TrackReconstruction} combine GPS data and DR, and the last three could be used with any tracking data.

\subsubsection*{Improving location estimation from GLS data}

\Rpkg{kftrack}, \Rpkg{kfsst} and \Rpkg{ukfsst} were developed by the same team of \Rpkg{trackit}, described in section \emph{Pre-processing} above. As \Rpkg{trackit}, they are mainly focused on fish movement. \Rpkg{kftrack}, \Rpkg{ukfsst} and \Rpkg{kfsst} use already estimated positions, either by the threshold method or given by the provider, and improve those estimations using a 2-dimensional random walk model \citep{Sibert2003}. Because of the generality of this modeling framework, \Rpkg{kftrack} could actually be used for any tracking data. In addition to the random walk model, \Rpkg{kfsst} includes SST as a covariate in the model \citep{Nielsen2006}, but it has been superseded by \Rpkg{ukfsst}, which implements an optimized parameter estimation. For that reason, we consider \Rpkg{kfsst} and \Rpkg{ukfsst} as one package.

\Rpkg{HMMoce}, also adapted to fish movement and working with already estimated/provided locations, uses HMMs (like \Rpkg{TwilightFree}) and incorporates depth-temperature profiles and SST as covariates in the observed model \citep{Braun2017}. 

\subsubsection*{Improving location estimation from PTT data}

\Rpkg{bsam} estimates locations by fitting Bayesian state-space models to the data. They offer the possibility of accounting for different movement patterns using `switching models' or HMMs; if this is opted out, first-difference correlated random walk models (DCRWs) are used. It is possible to estimate some of the model parameters for each individual and others at the population level (see \cite{Jonsen2013,Jonsen2016} for more details). The \Rpkg{argosTrack} package fits several types of movement models to PTT data \citep{Albertsen2015}, such as correlated random walks (CRWs) in discrete and continuous versions, and Ornstein-Uhlenbeck (OU) models, using Laplace approximation via Template Model Builder. 

\subsubsection*{Combining dead-reckoning and GPS data}

DR is based on vectorial calculations, thus even small errors in speed and/or direction accumulate over time. This can be further compounded in the presence of passive movements caused by currents and drifts. Independent locations, typically collected by a GPS recording at lower frequency than the accelerometers and magnetometers, are required to correct for these errors. \Rpkg{TrackReconstruction} provides a function that, after computing DR, forces the estimated locations to go through the known GPS points via space transformation, which returns a path with good shape but with biased length and orientation. \Rpkg{BayesianAnimalTracker} does not assume GPS to give the `true locations'. Instead, it implements a Bayesian approach to correct for biases, assuming a Brownian Bridge prior and using GPS points and an already estimated DR path to obtain a posterior of the sequence of locations. The posterior mean can be used as an estimate of the track, and the posterior standard error provides a measure of uncertainty about the estimated path \citep{Liu2016}. 

In \cite{Bidder2015}, the speed component is expressed as a linear equation, where the values of the coefficients are corrected iteratively until the dead-reckoned paths and ground-truth positions (e.g. GPS data) match. They also propose computing a correction factor for the heading vector. This method allows for correcting within the DR procedure, but has not been implemented in any R package so far.

\subsubsection*{Modeling movement of general tracking data}

\Rpkg{crawl} reconstructs paths by fitting continuous-time CRW models (called CTCRWs) \citep{Johnson2008} to tracking data. Though it can be used for any tracking data, \Rpkg{crawl} can account for the accuracy classes of PTT data to model the error associated with locations. \Rpkg{ctmcmove} fits a functional movement model \citep{Buderman2016} to the data and a set of probable true paths can be generated. \Rpkg{ctmm} fits several continuous movement models such as Brownian motion and OU-based models, selects the best models via AIC and allows for prediction (thus path reconstruction) with the selected model. 

\subsection*{Behavioral pattern identification}

Another common goal in movement ecology is to get a proxy of the individual's behavior through the observed movement patterns, based on either the locations themselves or second/third order variables such as distance, speed or turning angles. Covariates, mainly related to the environment, are frequently used for behavioral pattern identification. 

We classify the methods in this section as: 1) non-sequential classification or clustering techniques, where each fix in the track is classified as a given type of behavior, independently of the classification of the preceding or following fixes (i.e. independently of the temporal sequence); 2) segmentation methods, which identify change in behavior in time series of movement patterns to cut them into several segments; and 3) hidden Markov models, centered upon a hidden state Markovian process (representing the sequence of non-observed behaviors) that conditions the observed movement patterns \citep{Langrock2012}.

\subsubsection*{Non-sequential classification or clustering techniques}

\Rpkg{EMbC} (2015, CRAN, active) \citep{REMbC} implements the Expectation-maximization binary clustering method \citep{Garriga2016}. \Rpkg{m2b} (2017, CRAN, inactive) \citep{Rm2b} implements a random forest (a wrapper for the \Rpkg{randomForest} \citep{Liaw2002} package functions) to classify behaviors using a supervised training dataset, thus a dataset of both tracking data and known behaviors is needed to train the model.

\subsubsection*{Segmentation methods}

\Rpkg{adehabitatLT}, \Rpkg{bcpa} (2013, CRAN, inactive) \citep{Rbcpa}, \Rpkg{segclust2d}, \Rpkg{marcher} (2017, CRAN, active) \citep{Rmarcher} and \Rpkg{migrateR} (2016, GitHub, active) \citep{RmigrateR} implement segmentation methods. \Rpkg{adehabitatLT} presents two of these methods: Gueguen \citep{Gueguen2001} and Lavielle \citep{Lavielle1999,Lavielle2005}. \Rpkg{bcpa} implements the behavioral change point analysis \citep{Gurarie2009}. \Rpkg{segclust2d} implements a bivariate extension of Lavielle and is also described as an extension of \cite{Picard2007} by its authors, but there was no documentation on the method by the time of the review. Both \Rpkg{marcher} and \Rpkg{migrateR} are suited for analysis of migratory behavior. \Rpkg{marcher} enables the mechanistic range shift analysis method \citep{Gurarie2017} that identifies changes in locations of focal ranges, so migration and resident behaviors can be distinguished. The ranging models available in the package can take into account autocorrelation in location and in velocity. \Rpkg{migrateR} uses net displacement models to identify migratory, residency and nomadic behavior \citep{Spitz2017}. The models can incorporate factors such as elevation, sensitivity to starting date in the series, minimum time out of residence zone, among other features.  

\subsubsection*{Hidden Markov models} 

In this category we consider standard HMMs as well as more complex versions of these models; e.g. adding hierarchical structures, a second observation process for locations (state-space modeling), covariates affecting different components in the model, autoregressive processes or a spatial covariance structure. \Rpkg{bsam}, \Rpkg{lsmnsd} (2016, GitHub, active) \citep{Rlsmnsd}, \Rpkg{moveHMM} and \Rpkg{momentuHMM} implement methods that fall in the HMM category. \Rpkg{bsam}, for PTT data, implements Bayesian state-space models as described in section \emph{Path reconstruction} above, and may incorporate a layer of two switching states into the model: one state representing directed fast movement, and the other representing relatively undirected slow movement \citep{Jonsen2013}. \Rpkg{lsmnsd} use an HMM approach were the observed variable is net squared displacement and its mixture model distribution is conditioned on three hidden states that would correspond to two encamped and one exploratory mode \citep{Bastille-Rousseau2016}; the time spent in each mode and the transition probabilities are used to classify the track as migration, dispersal, nomadic or sedentary.

\Rpkg{moveHMM} and \Rpkg{momentuHMM} are not restricted to two or three states. \Rpkg{moveHMM} implements HMMs incorporating covariates and allowing for state sequence reconstruction, i.e. sequences of the behavioral proxies, via the Viterbi algorithm. In \Rpkg{moveHMM}, the variables modeled in the observed process are step length and turning angles, or two variables that statistically behave as step length and turning angles. \Rpkg{momentuHMM} implements generalized Hidden Markov models \citep{McClintock2012} with great flexibility for the choice of observed variables and their probability distributions, and covariate incorporation in the models. Since HMMs require regular time steps, \Rpkg{momentuHMM} offers a multiple imputation method \citep{McClintock2017}: it fits a CTCRW (from \Rpkg{crawl}) to the data obtaining regular time-step realizations and then fits an HMM to those realizations; all of this is done multiple times. Even if the data classes and model formulation in the package differ from \Rpkg{moveHMM}, many of the HMM-related functions are based on \Rpkg{moveHMM}. \Rpkg{moveHMM} is more user-friendly than \Rpkg{momentuHMM}, but \Rpkg{momentuHMM} offers greater modeling possibilities. 

\subsection*{Space and habitat use characterization}

Multiple packages implement functions to help answer questions related to where animals spend their time and what role environmental conditions play in movement or space-use decisions, which are typically split into two categories: home range calculation and habitat selection.

\subsubsection*{Home range}

Several packages allow the estimation of home ranges: \Rpkg{adehabitatHR}, \Rpkg{amt}, \Rpkg{BBMM} (2010, CRAN, inactive) \citep{RBBMM}, \Rpkg{ctmm}, \Rpkg{mkde} (2014, CRAN, inactive) \citep{Rmkde}, \Rpkg{MovementAnalysis}, \Rpkg{move}, \Rpkg{rhr} and \Rpkg{T-LoCoH}. They provide a variety of methods, from simple Minimum convex polygons (MCP) \citep{Mohr1947} to more complex probabilistic Utilization distributions (UD) \citep{VanWinkle1975}, potentially accounting for the temporal autocorrelation in tracking data, as detailed below.

\begin{itemize}
        \item \Rpkg{adehabitatHR} contains a comprehensive list of methods to estimate home ranges: convex hull methods like MCP, clustering techniques, Local convex hulls (LoCoH) \citep{Getz2007} and the characteristic hull method \cite{Downs2009}; UD methods like kernel home ranges, also with the modification from \cite{Benhamou2010} to account for boundaries, and methods to account for temporal autocorrelation between locations (Brownian bridge kernel method) \citep{Bullard1991}; biased random bridge kernel method also known as movement-based kernel estimation \citep{Benhamou2010, Benhamou2011}; and product-kernel algorithm, \cite{Horne2007}.
        \item \Rpkg{amt} also allows the estimation of home ranges using three common approaches not based on movement (MCP, LoCoh, and kernel UD), as well as movement-based UDs from fitted Step Selection Functions \citep[SSFs,][see below]{Fortin2005}. 
        \item \Rpkg{rhr} \citep{Signer2015} provides a graphical user interface to estimate home ranges using several non-movement based methods, such as parametric home ranges, MCP, kernel UD, or local convex hulls, as well as the Brownian Bridge kernel method (as a wrapper to the \Rpkg{adehabitatHR} function). Complementary analyses include time to statistical independence, site fidelity test (against random permutation of step lengths and angles), among others.
        \item \Rpkg{T-LoCoH} is focused on constructing home-range hulls \citep{Lyons2013}. A time-scale distance metric and a set of different nearest-neighbor criteria are available to choose which points to consider in a same hull. Hull metrics for space use, such as number of revisitations (repeated visits of an individual to the same hull) and their durations are also computed. Although the package was originally implemented for GPS data, it can be used for tracking data in general. 
        \item \Rpkg{BBMM}, \Rpkg{movementAnalysis} and \Rpkg{mkde} use Brownian bridge movement models to obtain UDs. \Rpkg{mkde} allows for a 3D extension of the Brownian bridges \citep{Tracey2014}.
        \item \Rpkg{move}, in turn, calculates UDs of tracking data via dynamic Brownian Bridge modeling \citep{Kranstauber2012} or uses MCP for home range estimation; for the latter, it imports functions from \Rpkg{adehabitatHR}.
        \item \Rpkg{ctmm} fits several candidate continuous-time movement models via a variogram regression approach \citep{Fleming2014}, which can account for spatial autocorrelation in locations and periodicity in space use \citep{Peron2016}. UDs are computed via an autocorrelated kernel estimator, where the autocorrelation term comes from the movement model previously fitted \citep{Fleming2015}. 
\end{itemize}

\subsubsection*{Habitat use}

The role of habitat features on animal space use, or habitat selection, can be investigated with any of the following four packages. 

\begin{itemize}
        \item \Rpkg{hab} (2015, GitHub, inactive) \citep{Rhab} enhances several utility functions of \Rpkg{adehabitatHS} \citep{Calenge2006}, \Rpkg{adehabitatHR} and \Rpkg{adehabitatLT}, and provides core functions to prepare, fit and evaluate SSFs while relying on \Rpkg{adehabitatLT} classes to handle trajectories. SSFs essentially investigate habitat selection along the trajectory, by comparing habitat features at observed step locations with those at alternative random steps taken from the same starting point \citep{Thurfjell2014}.
        \item \Rpkg{amt} contains functions and wrappers to streamline the process of fitting SSFs from pairs of coordinates defining locations, to the conditional logistic regression model. It also allows fitting of integrated step selection functions (iSSFs), in which both movement behavior and resource selection are modeled, and the role of environmental variables on each of these processes is investigated \citep{Avgar2016}. 
        \item In \Rpkg{ctmcmove}, the role of habitat features is investigated through a generalized linear model framework, for which these features are rasterized, and the animal track is first imputed via functional movement modeling and then discretized in a gridded space (more details in \cite{Hanks2015}). 
\end{itemize}

\subsubsection*{Non-conventional approaches for space use}

Other non-conventional approaches for investigating space use from tracking data can be found in \Rpkg{moveNT} (2017, GitHub, active) \citep{RmoveNT}, \Rpkg{recurse} (2017, CRAN, active) \citep{Bracis2018}, \Rpkg{rsMove}, \Rpkg{feedr} and \Rpkg{VTrack}. 

\begin{itemize}
        \item \Rpkg{moveNT} tackles space use analysis via network graph theory \citep{Bastille2018}. The procedure could be summarized as follows: 1) tracking data is represented over a gridded map and the number of transitions between pixels are counted; 2) the adjacency matrix, i.e. the counts of transitions, are then used to compute some network metrics at the pixel level; 3) a Gaussian mixture model is fitted to one of the metrics (user choice) to cluster values in two groups potentially representing patches and interpatch movement.
        \item \Rpkg{rsMove} implements a procedure to identify feeding sites from tracking data as a function of environmental variables (remote sensing data). It uses a random forest classification model from the \Rpkg{caret} package \citep{Rcaret}; however, there is no information about how to fix the parameters of the model, so users should go through the documentation of \Rpkg{caret} to understand and calibrate the model. An application of the method can be found in \cite{Remelgado2017}, but the parametrization is not described in the manuscript.
        \item \Rpkg{recurse} aims at computing number of revisitations to pre-defined areas and their duration. These areas can be defined by the user by entering their center of gravity (by default, the fixes in the track) and a radius. The vignette gives important criteria to use the functions and interpret the results, though there are no citations of scientific publications.
        \Rpkg{feedr} and \Rpkg{VTrack}, for radio and acoustic telemetry data, respectively, provide statistics on animal visits to given logger stations/receivers. 
\end{itemize}

\subsection*{Trajectory simulation} 

Simulating trajectories can be useful to test hypotheses concerning movement, by comparing the patterns of simulated movement from several alternative theoretical models, or the patterns in the simulated movement to those of real observed tracks. In addition, simulation allows the quantification of estimator uncertainty by parametric bootstrapping (e.g. \cite{Michelot2016}). As with other types of data analysis, simulations highly depend on the model used by the researcher. 
The tracking packages implement trajectory simulation mainly based on Hidden Markov models, correlated random walks, Brownian motions, L\'evy walks or Ornstein-Uhlenbeck processes. 

Packages that allow simulation of trajectories from movement models fitted to tracking data (i.e. parameters are estimated by the models) are \Rpkg{moveHMM}, \Rpkg{momentuHMM} (HMMs), \Rpkg{bsam} (DCRWs), \Rpkg{crawl} (CTCRWs), \Rpkg{argosTrack} (discrete and continuous CRWs, and OU processes) and \Rpkg{ctmm} (several continuous time movement models). These packages have been described in previous sections, and the simulations are presented as additional features after model fitting in their documentation. Another package for model fitting and simulation is \Rpkg{smam} (2013, CRAN, inactive) \citep{Pozdnyakov2018,Pozdnyakov2014,Pozdnyakov2017,Yan2014,Rsmam}. It can fit and simulate two types of movement models: Brownian motions with measurement error \citep{Pozdnyakov2014} and moving-resting processes with Brownian motion for the moving stage \citep{Yan2014}.

Other packages implement simulation functions when there is no previous model fitting to tracking data (i.e. movement parameters are known or simulations concern hypothetical mobile organisms). \Rpkg{adehabitatLT} proposes trajectory simulation using Brownian motion-based models, L\'evy walks, CRWs and bivariate OU motion. \Rpkg{trajr} allows for CRWs, directed random walks (direction is equal to a constant plus a small noise), Brownian motion and L\'evy walks. \Rpkg{moveNT} enables simulation of movement within and between patches. Movement within patches can follow an OU process (wrapping functions from \Rpkg{adehabitatLT}) or a two-states movement model (wrapping functions from \Rpkg{moveHMM}). Movement between patches is simulated via a Brownian bridge movement model (from \Rpkg{adehabitatLT}). 

\Rpkg{SiMRiv} (2016, CRAN, active) \citep{RSiMRiv} is another package created for simulation and it can take into account environmental constraints. It allows simulating random walks, correlated random walks, multi-state movement and constraining the area by an environmental resistance variable---defined by the user---that conditions the direction of the movement. The available documentation gives a detailed explanation of the simulation process. 

\subsection*{Other analyses of tracking data}

\subsubsection*{Interactions}

Interactions between individuals can be assessed using metrics from \Rpkg{wildlifeDI} (2014, CRAN, active) \citep{RwildlifeDI}, which quantifies the dynamic interaction between two tracks of distinct individuals through several metrics (see \cite{Long2014} for details). The package relies on `ltraj' objects (\Rpkg{adehabitatLT} data class for trajectories). Other packages that include functions investigating interaction are \Rpkg{TrajDataMining} and \Rpkg{movementAnalysis}: \Rpkg{TrajDataMining} can identify potential partners based on distance and time thresholds fixed by the user and \Rpkg{movementAnalysis} computes the expected duration of encounters at each location for every pair of IDs, based on a Brownian Bridge movement model fitted to the tracking data. 

\subsubsection*{Movement similarity}

\Rpkg{SimilarityMeasures} (2015, CRAN, inactive) \citep{RSimilarityMeasures} assesses similarity between trajectories using metrics such as the longest common subsequence (LCSS), Fr\'echet distance, edit distance and dynamic time warping (DTW). \cite{Magdy2015} provides a brief review on trajectory similarity measures. \Rpkg{trajectories} also computes the Fr\'echet distance for two trajectories. 

\subsubsection*{Population size}

\Rpkg{caribou} (2011, CRAN, inactive) \citep{Rcaribou} was specifically created to estimate population size from Caribou tracking data, but can also be used for wildlife populations with similar home-range behavior. The methods implemented here are described in \cite{Rivest1998}. The user needs to specify parameters concerning the size of each detected group, the number of collars in each of these groups and the detection model to use. 

\subsubsection*{Inferring environmental conditions}

Using tracking data to infer an environmental variable is the objective of \Rpkg{moveWindSpeed} (2016, CRAN, active) \citep{RmoveWindSpeed}. It uses avian tracking data to estimate wind speed via a maximum likelihood approach \citep{Weinzierl2016}. The estimation is only performed for segments where the bird is circling in a thermal, so a function in the package identifies those segments. Speed is modeled as a mean with an autocorrelated drift. 

\subsubsection*{Database management}

Finally, \Rpkg{rpostgisLT} (2016, CRAN, active) \citep{RrpostgisLT} handles database management for trajectory data by integrating R and the `PostgreSQL/PostGIS' database system. The package relies on \Rpkg{adehabitatLT}, and allows users to seamlessly transfer `ltraj' objects from R to the database, and vice-versa, using the corresponding `pgtraj' data structure in the database.

\subsection*{Analysis of biologging but not tracking data}

Time-depth recorders (TDRs) collect data on depth, velocity and other parameters as animals move through the water. These biologging data by themselves do not allow obtaining tracking data $(x,y,t)$ and thus comparable analyses to the ones presented above, however we briefly describe the R packages that could be used to analyze TDR and accelerometer data. \Rpkg{diveMove} \citep{Luque2007} and \Rpkg{rbl}, the latter also for accelerometer data, are the two packages implementing TDR data analysis. \Rpkg{diveMove} contains functions to identify wet and dry periods in the series, calibrate depth and speed sensor readings, identify individual dives and their phases, summarize statistics per dive and plot the data. With \Rpkg{rbl}, accelerometry data are used for identifying prey catch attempts \citep{Viviant2010} and swimming effort from frequency and magnitude of tail movement \citep{Bras2016}. Other functions allow the extraction of summary statistics from dives (e.g. maximum depth), fitting broken stick models (i.e. piecewise linear regression) to dive series and identifying dive phases. 

Accelerometry data are also used in human studies, primarily to assess levels of physical activity. Six R packages focus on the analysis of human accelerometry data, mainly to describe periodicity and levels of activity. \Rpkg{accelerometry} \citep{Raccelerometry}, \Rpkg{GGIR} \citep{vanHees2014,vanHees2015,RGGIR} and \Rpkg{PhysicalActivity} \citep{RPhysicalActivity} identify wear and non-wear time of the accelerometers. \Rpkg{nparACT} computes descriptive statistics such as interdaily stability, intradaily variability and relative amplitude of activity \citep{Blume2016}. \Rpkg{acc} \citep{Racc}, \Rpkg{GGIR} and \Rpkg{pawacc} \citep{Geraci2012,Rpawacc} classify wear data into different levels of activity (e.g. sedentary, moderate and vigorous) using thresholds given by the user, and offer some functions for visual representation of the data and descriptive statistics on the types of activities. Additionally, \Rpkg{acc} allows for activity simulation via Hidden Markov modeling.

\section*{Packages documentation}

Documentation in the form of manuals, vignettes (long-form documentation), tutorials or published articles is key to guide the use of a package's features, especially if the package contains a large number of functions and tools. Without proper user testing and peer editing, package documentation can lead to large gaps of understanding and limited usefulness of the package. If functions and workflows are not explicitly defined, a package's capacity to help users is undermined. Vignettes can act as road maps for the user, and published articles pertaining to the package help provide context and guidance on the internal workings of functions. Moreover, since packages make specific methods available for R users, the documentation should not only explain how to use the packages but also describe or provide references for the methods. 

To assess package documentation, an online survey was conducted between August and October 2018. The survey got Institutional Review Board exemption (IRB201802319). Questions in the survey regarded helpfulness of package documentation and the frequency of package use; it was completed by 225 people. The exact formulation of each question in the survey, detailed results and a discussion on the representativity of the survey are accessible in \url{https://doi.org/10.5281/zenodo.3483853}. 

Among 26 packages with at least 10 respondents, we identified 10 packages as having `adequate documentation', meaning that more than $75\%$ of the respondents expressed that the documentation was either good (allowing the user to do everything they wanted and needed to do with the package) or excellent (allowing users to do even more than what they initially planned because of the excellent quality of the information). These are: \Rpkg{momentuHMM} ($93.8\%$), \Rpkg{moveHMM} ($89.5\%$), \Rpkg{adehabitatLT} ($88.6\%$), \Rpkg{adehabitatHR} ($83.2\%$), \Rpkg{EMbC} ($81.8\%$), \Rpkg{wildlifeDI} ($81.3\%$), \Rpkg{ctmm} ($80.0\%$), \Rpkg{GeoLight} ($77.8\%$), \Rpkg{move} ($76.6\%$) and \Rpkg{recurse} ($76.5\%$) (see Fig. $3$). From this group of packages, \Rpkg{move} offers manuals and vignettes, while all the others offer in addition scientific articles centered on the package.

The results of this survey should be used by package developers as guidance to decide on whether to improve the documentation of their packages so more researchers can use them. 

\section*{Links between the packages}

We analyzed the links between tracking packages. If a package needs functions that have already been created by another package, the developer(s) can use those functions by declaring this dependency in the description file of the package under `Depends on', `Imports' or `Linking to' categories. Theoretically there are some differences between the three, but in practice developers mix those groups, so we consider them as part of the same concept: dependency. A package can also suggest using other packages; for instance, a package focused on data analysis can recommend, in the case data have to be cleaned first, the use of a package that allows post-processing. 

Developers usually define their own data classes for their packages. A data class allows them to pre-define the minimum requirements that data should have (e.g. dimensions, variables) and guarantee that the functions in the package will work if the data are in the pre-defined format. Similarly, if a package uses functions from other packages or the developer wants to facilitate the use of other packages along with their own, the latter should also provide coercion methods, i.e. functions that allow compatibility with data classes from these other packages. 

The dependency and suggestion information (collected in August 2018) was used for a graph analysis of package links (Fig. $4$). Thirty-nine packages in total showed some level of connections among them ($30$ in the form of one large group and three other small groups), while $19$ ($32\%$) of the packages worked in isolation. \Rpkg{adehabitatLT} and \Rpkg{move} were the most suggested/depended-on packages with 14 and 8 links to them, respectively (8 and 2, respectively, were dependencies). Indeed, many packages use functions compatible with the `ltraj' data class from \Rpkg{adehabitatLT}, and some others with the `move' class from \Rpkg{move}. \Rpkg{amt} suggests more packages than any other (6), and it provides coercion methods for data classes from the packages it suggests. 

\section*{Discussion}

As the quantity and diversity of biologging data increases, so does the need for suitable statistical techniques and software resources. These tools are essential to convert collected data into ecologically meaningful measures and analyze outputs to test hypotheses. Through a systematic search we identified 58 R packages aimed at processing or analyzing tracking data. The packages offer tools for data processing, visualization, computation of statistics for track description, path reconstruction, behavioral pattern identification, space use characterization and trajectory simulation, among others. 
All the stages of the movement ecology workflow are covered by the reviewed packages. In some cases, there is even function overlapping, with more than one package implementing the same type of analysis with the same or very similar approaches (e.g. \Rpkg{animalTrack} and \Rpkg{TrackReconstruction} for DR, \Rpkg{BBMM}, \Rpkg{MovementAnalysis} and \Rpkg{mkde} for Brownian bridge movement models). 
A type of analysis that was poorly covered was collective motion: mainly \Rpkg{wildlifeDI} and, to a lesser degree, \Rpkg{TrajDataMining} and \Rpkg{movementAnalysis} allow computing descriptive metrics on encounters between individuals, periods of proximity or other metrics of interaction. The lack of R functions to analyze collective movement beyond descriptive statistics is most likely a reflection of the early stages of this field regarding the use of tracking data; collective behavior has mostly relied on controlled laboratory-based studies and theoretical models \citep{Westley2018}.
Overall, the review highlighted the abundance of analytical tools available, but also identified a need to improve visibility and accessibility (i.e. documentation) to existing packages more than developing new packages.     

\subsection*{Integration over proliferation}

Transparency in science is facilitated by the sharing of data and analytical tools, including code. This has resulted in a general tendency in the scientific community to convert functions into publicly available packages. In movement ecology, this has translated into a proliferation of R packages dealing with tracking data, many of them, isolated from all other packages (Fig. $4$) despite having similar goals and methods. While a large number of packages reflects that the field is active and that codes for several types of analyses are available for the community, such independent proliferation of packages makes it hard to maintain an overview of their functionality and availability. Here we presented a list of 58 packages but the number is expected to keep increasing steadily, associated with an increased possibility of unnecessary redundancy and disconnection between the packages. Due to the already overwhelming number of tracking packages, we suggest developers only create new packages in the future when they represent a new and complementary contribution to the scientific and programming community. 

While package necessity is not assessed through any repository, there is a peer-review process available for packages through rOpenSci, a non-profit initiative founded in 2011 with the goal of making scientific data more retrievable and reproducible (\url{http://ropensci.org}). Packages submitted to rOpenSci are reviewed by two independent reviewers for readability, usability, usefulness and redundancy. The rOpenSci community checks that developers adhere to coding `best practices' such as unit testing (i.e. testing if individual units of code work correctly), continuous integration (i.e. all changes made by developers are immediately tested and reported when added to the mainline code base), minimizing code duplication, and strong documentation. This open review process improves packages as it helps developers strengthen their package and coding, while gaining additional technical support from rOpenSci's volunteer staff. In addition, a couple of journals have partnered with rOpenSci: the Journal of Open Source Software (JOSS, \url{http://joss.theoj.org}) and Methods in Ecology and Evolution (MEE, \url{methodsinecologyandevolution.org}). JOSS is an open access journal for research software packages that adheres to similar standards as rOpenSci, and, if the submitted package have already been accepted to rOpenSci, they can be submitted for fast-track publication at JOSS, in which JOSS editors may evaluate based on rOpenSci's reviews alone. MEE can publish articles on new R packages and gives authors the option of a joint rOpenSci-MEE review in which the package is reviewed by rOpenSci, followed by fast-tracked review of the manuscript by MEE. The R Journal (\url{https://journal.r-project.org/}) and, for packages concerning statistical analysis, the Journal of Statistical Software (\url{https://www.jstatsoft.org/}), are other choices of journal, that adhere to similar standards as rOpenSci. 

\subsection*{Recommendations}

This work is not intended to tell ecologists exactly which packages to use, but to provide them an exhaustive catalog of tracking packages, a description of their functions and show the similarities and differences between them. We suggest researchers use packages with good documentation, that are actively maintained and that have a large number of users. Good documentation facilitates the initial use of a package. A regularly maintained package means that there is a person or team behind it, and that, when an error arises in the package, it will likely be fixed rapidly and a new version will be available. A package that has a large number of users means greater opportunity to 1) identify bugs in the package, calling the attention of the maintainer for a rapid fix, and thus improving functionality, and to 2) obtain additional guidance on package use from other users. Regarding the methods available in the packages, while we previously stated the importance of describing them and citing references, it is the responsibility of the researchers to select and apply a method if they correctly understand it, and not only because it is available in a package. Also, with a critical use of packages, researchers should feel encouraged to report bugs when they see them, to contribute to their improvement.

When developers are working on new packages, we recommend they submit to rOpenSci and consider the following criteria:
\begin{itemize}
        \item Contribution: Does your package fill a gap or need? Does a function within the package perform a novel task that does not already exist in another published package? Can those functions be instead added to an existing package? Developers should contemplate the possibility (and appropriateness) of contacting authors of existing and actively maintained packages to incorporate new functions. We also suggest the authors of existing packages to be open to integration of new functions (and new collaborators) within their package.
        
        \item Data class coercion: Does the package handle commonly used data classes (e.g. spatial classes from sp or `ltraj' from \Rpkg{adehabitatLT}), so that it is compatible with the use of other packages? Since tracking packages deal with spatial data, most of them use georeferenced data classes. \Rpkg{sp} data classes \citep{Pebesma2005,Bivand2013} are the most popular spatial data classes ($40$ out of the $58$ packages use them). The recent \Rpkg{sf} package \citep{Pebesma2018} aims at providing a simpler and standard implementation of geographic objects in R; handling \Rpkg{sf} objects is as easy as handling non-spatial objects in R, and computationally more efficient than using \Rpkg{sp}. Only one package (\Rpkg{crawl}) was compatible with \Rpkg{sf} at the time our research was done. Because of its functionalities, we encourage developers to provide coercion methods to \Rpkg{sf}. Regarding data classes for trajectories, ltraj (from \Rpkg{adehabitatLT}) is one of the oldest and most used data classes, but others exist (e.g. trip from \Rpkg{trip} or Track from \Rpkg{trajectories}). Ideally, the community of tracking-package developers should unite to discuss the best data class for a trajectory, and once a consensus is reached, provide coercion methods to that class. 
        \item Documentation: Is the documentation clear, exhaustive on the functions, with methods description or references available? The latter is even more important if the package implements a new method of analysis. Worked examples and vignettes can enable researchers to navigate through the package and learn what it does more easily, minimizing the need for additional support.
        \item Maintenance: Who will maintain the package over time? Specific people are required to maintain the continuity of packages. Typically, lab PIs  or members of a working group/collaboration could take this role in view of the long term commitment. On CRAN, non-maintained packages are considered `orphaned' if they are not actively maintained and `archived' if they do not pass `R CMD check' anymore (\url{https://cran.r-project.org/src/contrib/Orphaned/README}). 
\end{itemize}

\section*{Conclusions}

This review has served as a road map of the tools implemented by the packages for data analysis in movement ecology. In recent years, programmers have responded to the need for advanced statistical tools to analyze movement data by developing at least 58 R tracking packages. However, we emphasize that increased accessibility and understanding of existing packages (in which documentation plays a fundamental role), and more integration for package development will help the advancement of research in this field, allowing researchers to continue to address novel and exciting questions.

\section*{Acknowledgments}

The authors (RJ, TAC, SCP, SCT and MB) were funded by a Human Frontier Science Program Young Investigator Grant (SeabirdSound - RGY0072/2017). We are grateful to Guillaume Bastille-Rousseau, Cl\'ement Calenge, Christen H. Fleming, Devin Johnson, Bart Kranstauber, Simeon Lisovski, Brett McClintock, Benjamin Merkel, Th\'eo Michelot, Anders Nielsen, Eldar Rakhimberdiev, Henry Scharf, Jakob Schwalb-Willmann, Takashiro Shimada, Derek Spitz, and Michael Sumner for enlightening discussions and additional information about their packages. 
Special thanks to Simon Benhamou for discussions about accelerometers, magnetometers and gyroscopes, and reviewing the corresponding sections of the manuscript. During review, constructive comments and suggestions of Johannes Signer and Luca Borger (associated editor) have significantly contributed to the quality of this work. 
We also thank the anonymous survey participants and everyone who suggested packages for this review.

\section*{Data accessibility}

The data collected on all the packages and used for the review and the anonymous data from the survey, as well as R codes and additional files to reproduce Figures 2--4 are available on Zenodo \url{https://doi.org/10.5281/zenodo.3483853}. \\

\section*{Authors' contributions}
RJ conceived the ideas of the review and reviewed the R packages. RJ and MB worked on the workflow for the manuscript. RJ and MEB gathered the information on the packages. RJ led the survey and analyzed the results; MEB and MB implemented it in an online platform. RJ, MEB and MB worked in the Zenodo repository. RJ led the writing of the manuscript. All authors contributed critically to the drafts and reponses to the reviewers, and gave final approval for publication.

\newpage

 \bibliographystyle{abbrv}

\newpage
\section*{Table captions}
\noindent \textbf{Table~1.} Summary of the functionality of the tracking packages.\\

\newpage

\begin{landscape}
        \singlespacing
        \begin{longtable}{>{\arraybackslash}p{1.905cm}p{2cm}p{5.2cm}p{1.7cm}p{7.7cm}}
                \hline
                Workflow stage & Categories & Method description & Data type & Package   \\ 
                \hline
                \multirow{2}{1.9cm}{Pre-processing} & & Threshold & GLS & \Rpkg{GeoLight}, \Rpkg{probGLS} \\
                & & Template-fitting & GLS & \Rpkg{FlightR}, \Rpkg{trackit}, \Rpkg{TripEstimation/SGAT} \\
                & & Twilight-free & GLS & \Rpkg{TwilightFree} \\
                & & Triangulation & Radio & \Rpkg{telemetr} \\
                & & Dead reckoning & Accel. + magnet. & \Rpkg{animalTrack}, \Rpkg{TrackReconstruction} \\
                \hline
                \multirow{2}{1.9cm}{Post-processing} & \multirow{2}{1.9cm}{Data cleaning} & Filter implausible locations & PTT & \Rpkg{argosfilter}, \Rpkg{SDLfilter} \\
                & & & GPS & \Rpkg{SDLfilter} \\
                & & \multirow{2}{5.2cm}{Remove duplicates / speed filter} & Any & \Rpkg{T-LoCoH}, \Rpkg{TrajDataMining}, \Rpkg{trip} \\ 
                & & & & \\
                \cmidrule{2-5}
                & \multirow{2}{1.9cm}{Data compression} & Rediscretization & Any & \Rpkg{adehabitatLT}, \Rpkg{trajectories}, \Rpkg{trajr} \\
                & & Interpolation & Any & \Rpkg{adehabitatLT}, \Rpkg{amt} , \Rpkg{trajectories}\\
                & & Douglas-Peucker & Any & \Rpkg{TrajDataMining}, \Rpkg{trajectories} \\
                & & Opening window & Any & \Rpkg{TrajDataMining} \\
                & & Savitzky-Golay & Any & \Rpkg{trajr} \\
                & & \multirow{2}{5.2cm}{Transform to pixels to link with remote sensing} & Any & \Rpkg{rsMove} \\
                & & & & \\
                \cmidrule{2-5}
                & \multirow{3}{1.9cm}{Metrics computation} & 2nd or 3rd order variables & Any &  \Rpkg{adehabitatLT}, \Rpkg{amt}, \Rpkg{bcpa}, \Rpkg{momentuHMM}, \Rpkg{move}, \Rpkg{moveHMM}, \Rpkg{rhr}, \Rpkg{segclust2d}, \Rpkg{trajectories}, \Rpkg{trajr}, \Rpkg{trip} \\
                & & & Radio & \Rpkg{feedr} \\
                & & & Acoustic & \Rpkg{VTrack} \\
                \hline
                Visualization & & Animations of tracks & Any & \Rpkg{anipaths}, \Rpkg{moveVis} \\
                \hline
                \multirow{2}{1.9cm}{Track description} & & Summary metrics & Any & \Rpkg{amt}, \Rpkg{movementAnalysis}, \Rpkg{trajr}, \Rpkg{marcher} \\
                & & & GPS & \Rpkg{trackeR} \\ 
                \hline
                \multirow{3}{1.9cm}{Path reconstruction} & & State-space models & GLS & \Rpkg{HMMoce}, \Rpkg{kftrack}, \Rpkg{ukfsst/kfsst} \\ 
                & & & PTT & \Rpkg{argosTrack}, \Rpkg{bsam} \\
                & & & Any & \Rpkg{crawl} \\
                & & Functional movement model & Any & \Rpkg{ctmcmove} \\
                & & \multirow{2}{5.2cm}{Continuous Markov chain in gridded space} & Any & \Rpkg{ctmm} \\
                & & & & \\
                & & Bayesian Brownian bridge model & GPS + DR path & \Rpkg{BayesianAnimalTracker} \\
                & & Transformation of the space & GPS + DR path & \Rpkg{TrackReconstruction} \\
                \hline
                \multirow{3}{1.9cm}{Behavioral pattern identification} & \multirow{2}{1.9cm}{Clustering techniques} & \multirow{2}{5.2cm}{Expectation-maximization binary clustering} & Any & \Rpkg{EMbC} \\
                & & & & \\
                & & Random forest & Any & \Rpkg{m2b} \\
                \cmidrule{2-5}
                & Segmentation & Gueguen and Lavielle & Any & \Rpkg{adehabitatLT} \\
                & & Extension of Lavielle & Any & \Rpkg{segclust2d} \\
                & & \multirow{2}{5.2cm}{Behavioral change point analysis} & Any & \Rpkg{bcpa} \\
                & & & & \\
                & & \multirow{2}{5.2cm}{Mechanistic range shift analysis} & Any & \Rpkg{marcher} \\
                & & & & \\
                & & Net displacement models & Any & \Rpkg{migrateR} \\
                \cmidrule{2-5}
                & \multirow{4}{2cm}{Hidden Markov based models} & \multirow{2}{5.2cm}{Bayesian state-space model with states} & PTT & \Rpkg{bsam} \\
                & & & & \\
                & & Hidden Markov models & Any & \Rpkg{lsmnsd}, \Rpkg{momentuHMM}, \Rpkg{moveHMM} \\
                & & & & \\
                \hline
                Space use & \multirow{3}{2cm}{Home range estimation} & Minimum convex polygon & Any & \Rpkg{adehabitatHR}, \Rpkg{amt}, \Rpkg{move}, \Rpkg{rhr} \\
                & & \multirow{2}{5.2cm}{Density kernel utilization distribution} & Any & \Rpkg{adehabitatHR}, \Rpkg{amt}, \Rpkg{rhr} \\
                & & & & \\
                & & \multirow{2}{5.2cm}{Movement-based utilization distribution} & Any & \Rpkg{adehabitatHR}, \Rpkg{amt}, \Rpkg{BBMM}, \Rpkg{ctmm}, \Rpkg{mkde}, \Rpkg{move}, \Rpkg{movementAnalysis}, \Rpkg{rhr} \\
                & & Local convex hull & Any & \Rpkg{adehabitatHR}, \Rpkg{amt}, \Rpkg{rhr}, \Rpkg{T-LoCoH} \\
                \cmidrule{2-5}
                & \multirow{2}{2cm}{Habitat use} & Step selection functions & Any & \Rpkg{amt}, \Rpkg{hab} \\
                & & Generalized linear models & Any & \Rpkg{ctmcmove} \\
                & \multirow{3}{1.9cm}{Non-conventional approaches} & (See text) & Radio & \Rpkg{feedr} \\
                & & & Acoustic & \Rpkg{VTrack} \\
                & & & Any & \Rpkg{moveNT}, \Rpkg{recurse}, \Rpkg{rsMove} \\
                \hline
                \multirow{2}{1.9cm}{Trajectory simulation} & & \multirow{2}{5.2cm}{Movement models fitted to data} & Any & \Rpkg{crawl}, \Rpkg{ctmm}, \Rpkg{momentuHMM}, \Rpkg{moveHMM}, \Rpkg{smam} \\
                & & & PTT & \Rpkg{argosTrack}, \Rpkg{bsam} \\
                \cmidrule{3-5}
                & & \multirow{2}{5.2cm}{Movement models with parameters defined by user} & Any & \Rpkg{adehabitatLT}, \Rpkg{moveNT}, \Rpkg{SiMRiv}, \Rpkg{trajr} \\
                & & & & \\
                \hline
                Other & Interactions & Dyad interaction metrics & Any & \Rpkg{wildlifeDI} \\
                & & \multirow{2}{5.2cm}{Distance and time thresholds} & Any & \Rpkg{movementAnalysis}, \Rpkg{TrajDataMining} \\
                & & & & \\
                \cmidrule{2-5}
                & \multirow{2}{1.9cm}{Movement similarity} & \multirow{2}{5.2cm}{Similarity measures between trajectories (e.g. Frechet)} & Any & \Rpkg{SimilarityMeasures}, \Rpkg{trajectories} \\
                & & & & \\
                \cmidrule{2-5}
                & \multirow{2}{1.9cm}{Population size} & \multirow{2}{5.2cm}{Stochastic model for abundance} & Radio & \Rpkg{caribou} \\
                & & & & \\
                \cmidrule{2-5}
                & Environment conditions & \multirow{2}{5.2cm}{Likelihood maximization of airspeed model} & Any & \Rpkg{moveWindSpeed} \\
                \cmidrule{2-5}
                & Database management & \multirow{2}{5.2cm}{Integrating R and PostgreSQL / PostGIS} & Any & \Rpkg{rpostgisLT} \\
                \hline
                \label{table:PurposeTable}
        \end{longtable}
        
\end{landscape}

\section*{Figure captions}

\noindent \textbf{Figure~1.} Workflow for data processing and analysis in movement ecology. Numbers in parenthesis are the number of packages dealing with each stage of the workflow. Some packages may correspond to more than one category, except for data visualization, where only packages created for that purpose are counted.\\

\noindent \textbf{Figure~2.} Number of packages per year of publication. Since the packages were reviewed between March and August 2018, this last year was incomplete and not included in the graph. \\

\noindent \textbf{Figure~3.} Packages with good and excellent documentation (survey results). Text color in green corresponds to packages with standard documentation only, blue is for packages with vignettes, and purple is for packages that also have peer-reviewed articles published. Only results for packages with at least 10 respondents are shown. \\

\noindent\textbf{Figure~4.} Network representation of the dependency and suggestion between tracking packages. The arrows go towards the package the others suggest (dashed arrows) or depend on (solid arrows). Bold font corresponds to active packages. The size of the circle is proportional to the number of packages that suggest or depend on this one.

\newpage
\begin{figure}[h!]
  \caption{}
  \label{Fig1}
  \begin{center}
    \includegraphics[width=\textwidth]{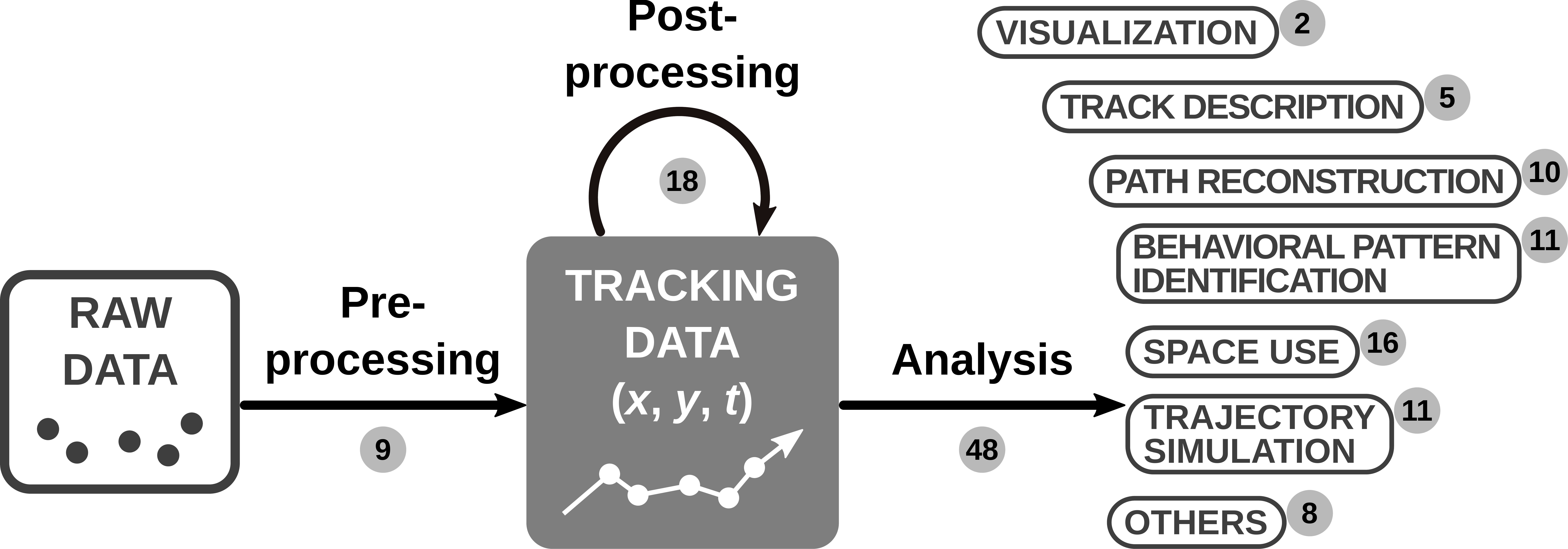}

  \end{center}
\end{figure}

\newpage
\begin{figure}[h!]
  \caption{}
  \label{Fig2}
  \begin{center}
    \includegraphics[width=0.75\textwidth]{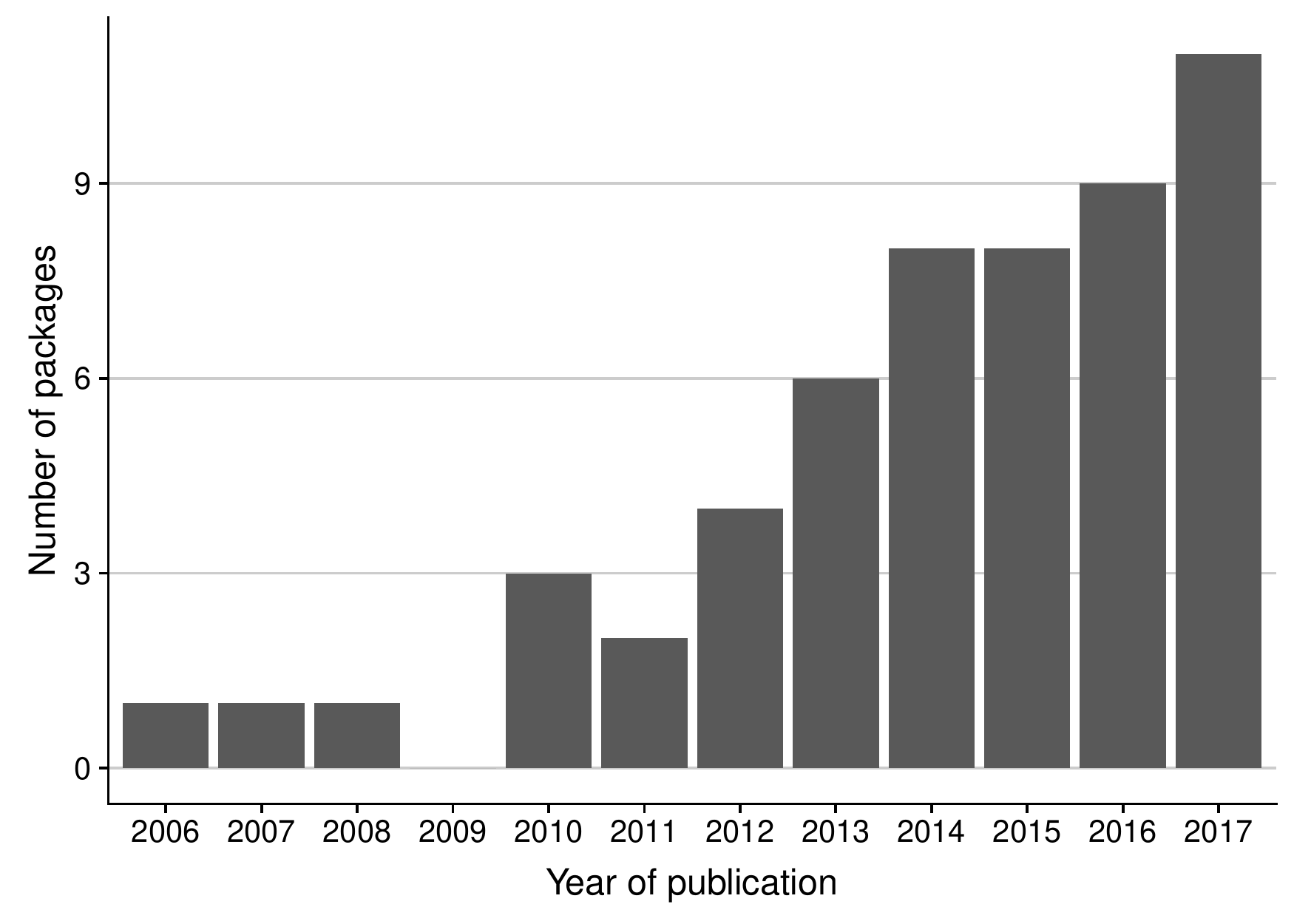}

  \end{center}
\end{figure}

\newpage
\begin{figure}[h!]
        \caption{}
        \label{Fig3}
        \begin{center}
                \includegraphics[width=\textwidth]{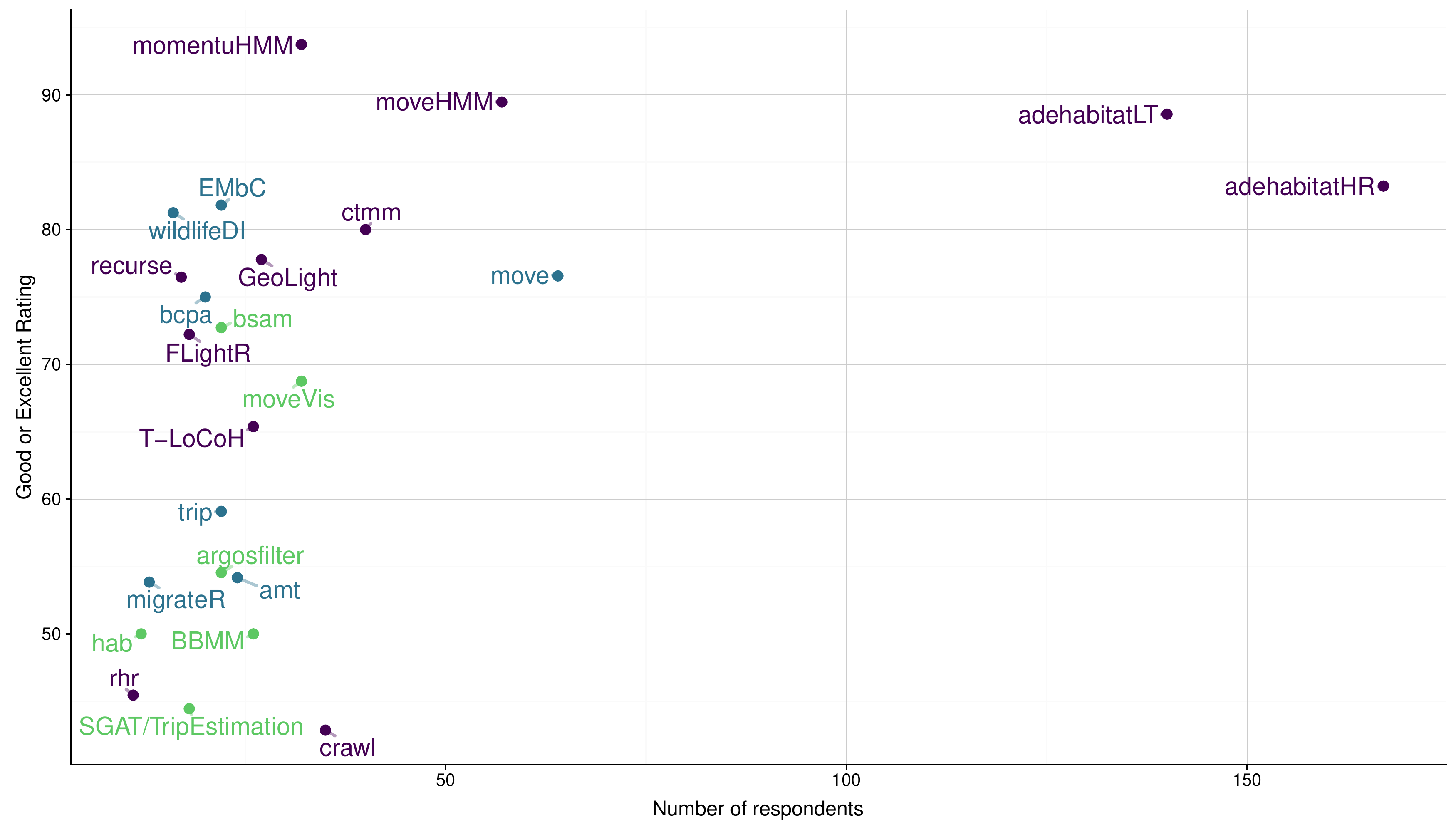}
        \end{center}
\end{figure}

\newpage
\begin{figure}[h!]
        \caption{}
        \label{Fig4}
        \begin{center}
                \includegraphics[width=1\textwidth]{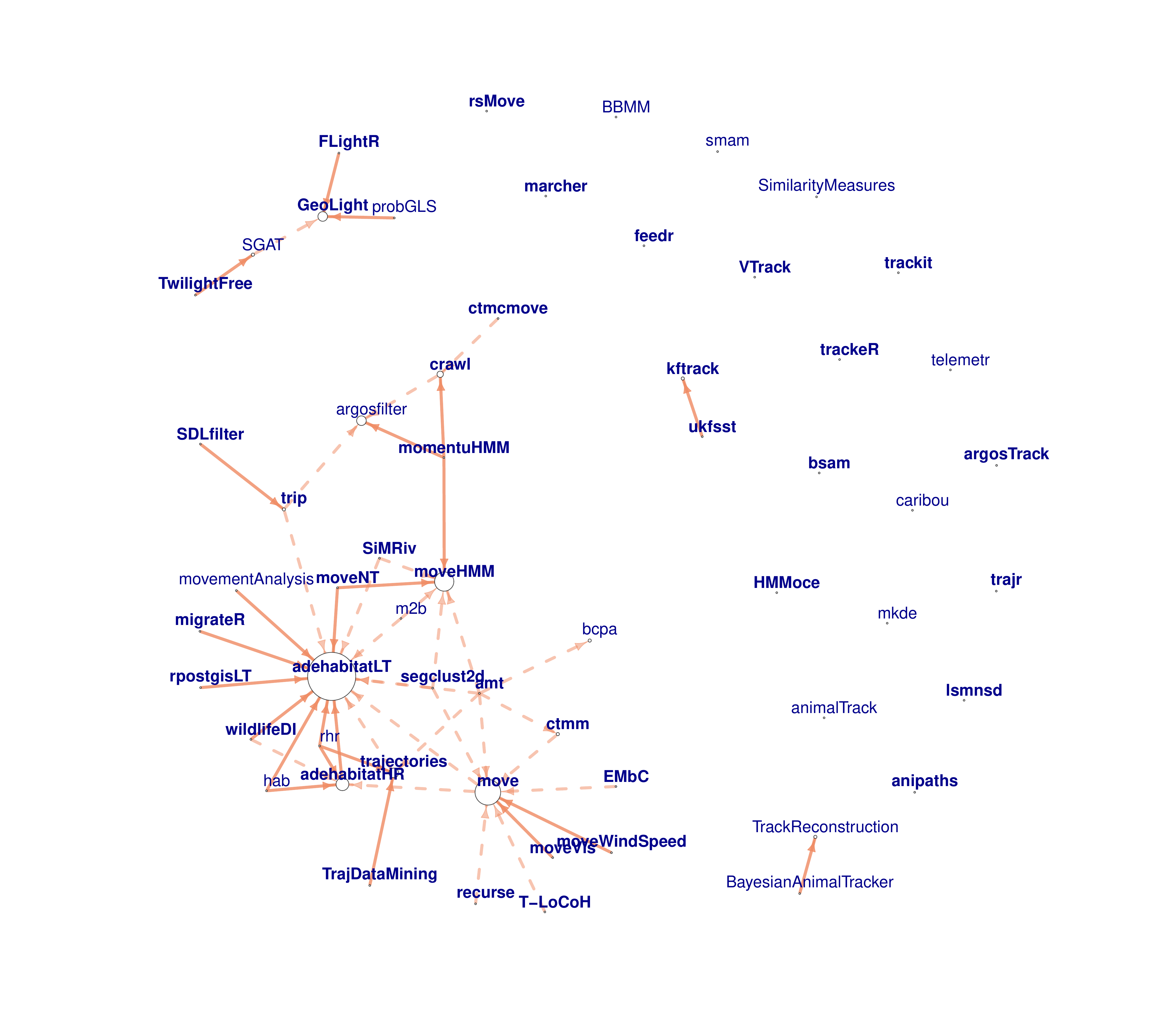}
        \end{center}
\end{figure}

\end{document}